\PassOptionsToPackage{unicode}{hyperref}
\PassOptionsToPackage{hyphens}{url}
\documentclass[
  a4paper,
]{article}
\usepackage{amsmath,amssymb}
\usepackage{lmodern}
\usepackage{iftex}
\ifPDFTeX
  \usepackage[T1]{fontenc}
  \usepackage[utf8]{inputenc}
  \usepackage{textcomp} 
\else 
  \usepackage{unicode-math}
  \defaultfontfeatures{Scale=MatchLowercase}
  \defaultfontfeatures[\rmfamily]{Ligatures=TeX,Scale=1}
\fi
\IfFileExists{upquote.sty}{\usepackage{upquote}}{}
\IfFileExists{microtype.sty}{
  \usepackage[]{microtype}
  \UseMicrotypeSet[protrusion]{basicmath} 
}{}
\makeatletter
\@ifundefined{KOMAClassName}{
  \IfFileExists{parskip.sty}{%
    \usepackage{parskip}
  }{
    \setlength{\parindent}{0pt}
    \setlength{\parskip}{6pt plus 2pt minus 1pt}}
}{
  \KOMAoptions{parskip=half}}
\makeatother
\usepackage{xcolor}
\IfFileExists{xurl.sty}{\usepackage{xurl}}{} 
\IfFileExists{bookmark.sty}{\usepackage{bookmark}}{\usepackage{hyperref}}
\hypersetup{
  pdftitle={Multilevel Quality Indicators (MQI)},
  hidelinks,
  pdfcreator={LaTeX via pandoc}}
\usepackage[margin=1in]{geometry}
\usepackage{longtable,booktabs,array}
\usepackage{calc} 
\usepackage{etoolbox}
\makeatletter
\patchcmd\longtable{\par}{\if@noskipsec\mbox{}\fi\par}{}{}
\makeatother
\IfFileExists{footnotehyper.sty}{\usepackage{footnotehyper}}{\usepackage{footnote}}
\makesavenoteenv{longtable}
\usepackage{graphicx}
\makeatletter
\def\maxwidth{\ifdim\Gin@nat@width>\linewidth\linewidth\else\Gin@nat@width\fi}
\def\maxheight{\ifdim\Gin@nat@height>\textheight\textheight\else\Gin@nat@height\fi}
\makeatother
\setkeys{Gin}{width=\maxwidth,height=\maxheight,keepaspectratio}
\makeatletter
\def\fps@figure{htbp}
\makeatother
\usepackage[normalem]{ulem}
\newlength{\cslhangindent}
\newlength{\csllabelwidth}
\newlength{\cslentryspacingunit} 
\newenvironment{CSLReferences}[2] 
 {
  \setlength{\parindent}{0pt}
  \ifodd #1
  \let\oldpar\par
  \def\par{\hangindent=\cslhangindent\oldpar}
  \fi
  \setlength{\parskip}{#2\cslentryspacingunit}
 }%
 {}
\usepackage{calc}

\newcommand{\CSLLeftMargin}[1]{\parbox[t]{\csllabelwidth}{#1}}
\newcommand{\CSLRightInline}[1]{\parbox[t]{\linewidth - \csllabelwidth}{#1}\break}

\ifLuaTeX
  \usepackage{selnolig}  
\fi

\title{Multilevel Quality Indicators (MQI)}
\usepackage{etoolbox}
\makeatletter
\providecommand{\subtitle}[1]{
  \apptocmd{\@title}{\par {\large #1 \par}}{}{}
}
\makeatother
\subtitle{Methodology and Monte Carlo evidence}
\author{Martin Roessler\(^a\), Claudia Schulte\(^a\), Uwe Repschläger\(^a\), Dagmar Hertle\(^a\), Danny Wende\(^a\)\\
\strut \\
\(^a\)BARMER Institut für Gesundheitssystemforschung (bifg)}
\date{}

\begin{document}
\maketitle

\begin{abstract}
\noindent \textbf{Background}: Quality indicators are frequently used to assess the performance of healthcare providers, in particular hospitals. Established approaches to the design of such indicators are subject to distortions due to indirect standardization and high variance of estimators. Indicators for geographical regions are rarely considered. \\

\noindent \textbf{Objectives}: To develop and evaluate a methodology of Multilevel Quality Indicators (MQI) for both healthcare providers and geographical regions.   \\

\noindent \textbf{Research Design}: We formally derived MQI from a statistical multilevel model, which may include characteristics of patients, providers, and regions. We used Monte Carlo simulation to assess the performance of MQI relative to established approaches based on the standardized mortality/morbidity ratio (SMR) and the risk-standardized mortality rate (RSMR). \\

\noindent \textbf{Measures}: Rank correlation between true provider/region effects and quality indicator estimates; shares of the 10\% best and 10\% worst providers identified by the quality indicators. \\

\noindent \textbf{Results}: The proposed MQI are 1) standardized hospital outcome rate (SHOR), 2) regional SHOR (RSHOR), and 3) regional standardized patient outcome rate (RSPOR). Monte Carlo simulations indicated that the SHOR provides substantially better estimates of provider performance than the SMR and RSMR in almost all scenarios. RSPOR was slightly more stable than the regional SMR. We also found that modeling of regional characteristics generally improves the adequacy of provider-level estimates.  \\

\noindent \textbf{Conclusions}: MQI methodology facilitates adequate and efficient estimation of quality indicators for both healthcare providers and geographical regions.
\end{abstract}

\hypertarget{introduction}{%
\section{Introduction}\label{introduction}}

Quality indicators are used to assess and compare the performance of healthcare providers, in particular hospitals.\textsuperscript{1--3} Over the last decades, numerous quality indicators have been designed for different patient groups and various outcomes. At the same time, there has been extensive research on the methodological foundations of quality measurement in healthcare. This research has highlighted several methodological issues and potentials for improvement. As briefly described in the following, key methodological challenges include the choice of an appropriate standardization approach and issues related to statistical efficiency. Moreover, while many indicators aim to measure quality at the level of healthcare providers, there is a lack of broader, policy-relevant indicators at the level of geographical regions.

\hypertarget{indirect-vs.-direct-standardization}{%
\subsection{Indirect vs.~direct standardization}\label{indirect-vs.-direct-standardization}}

Most quality indicators are adjusted for patient-specific risk factors to account for case mix differences between healthcare providers. While there is considerable methodological heterogeneity\textsuperscript{4--6}, the most common approach to risk adjustment in practical applications is based on the standardized mortality/morbidity ratio (SMR)\textsuperscript{7--9}. The SMR relates each healthcare provider's observed outcome rate (e.g., mortality, morbidity, or complication rate) to the outcome rate that would be expected based on the provider's case mix. The provider's expected outcome rate is usually derived from a logistic regression model.

Previous studies showed that measures of indirect standardization such as the SMR are subject to distortions that may induce misleading results.\textsuperscript{10--12} In principle, direct standardization overcomes the shortcomings of indirect standardization. However, conventional approaches to direct standardization require that all patient groups considered in risk adjustment were treated by all considered healthcare providers.\textsuperscript{13} This requirement is almost never fulfilled. A possible solution is direct standardization based on a (semi-)parametric statistical model.\textsuperscript{13--15} Despite its advantages, this approach is rarely used in practical applications.

\hypertarget{statistical-efficiency}{%
\subsection{Statistical efficiency}\label{statistical-efficiency}}

An essential assumption that underlies quality indicators is that the performance of healthcare providers has sufficiently strong, measurable influence on patient outcomes. The validity of this assumption has been questioned, particularly regarding quality indicators focusing on hospital mortality.\textsuperscript{16,17} From a statistical viewpoint, weak influence of provider performance on patient outcomes corresponds to a low signal-to-noise ratio. This low signal-to-noise ratio results in a high variance of estimators of provider performance. Consequently, empirical estimates may not facilitate reliable assessment and comparison of healthcare providers.\textsuperscript{18} Since the variance of estimators is, in general, inversely related to sample size, this issue is particularly relevant for the assessment of healthcare providers treating a small number of patients.

A promising approach that has the potential to address the problem of low signal-to-noise ratios effectively has recently been proposed in the methodological literature.\textsuperscript{15,19} The basic idea of this approach is to exploit information on healthcare provider characteristics that are related to quality of care. Together with patient-specific risk factors, those characteristics are included in a statistical model for the considered patient outcome. For instance, the case volumes of hospitals may be included in a model for hospital mortality if there is reason to expect a volume-outcome relationship. The estimation results of this model may then be used to derive performance estimates for single providers. This procedure deviates from conventional approaches to the design of quality indicators by using information from the \emph{whole sample} to characterize specific healthcare providers. The resulting gains in precision of estimates have particularly been demonstrated for small-volume hospitals.\textsuperscript{15}

\hypertarget{provider-level-and-area-level-indicators}{%
\subsection{Provider-level and area-level indicators}\label{provider-level-and-area-level-indicators}}

Generally, the task of healthcare providers is to improve or preserve the health status of their patients. Consequently, the provider level is the natural focal point of quality indicators. While numerous quality indicators have been developed to assess provider performance, only few indicators aim to measure healthcare quality within geographical regions.\textsuperscript{20--22} However, such area-level indicators can be valuable for patients, healthcare professionals, and policy makers as they may help to identify regions that lack behind in terms of population health and healthcare quality. Such insights may have important policy implications. Hence, developing suitable area-level indicators is an important but neglected task in healthcare quality measurement.

\hypertarget{objectives}{%
\subsection{Objectives}\label{objectives}}

Against the background outlined above, this paper proposes an approach to the design of Multilevel Quality Indicators (MQI). The MQI methodology derives quality indicators for both healthcare providers and geographical regions based on a single statistical model. This model may include characteristics of providers and regions to increase the precision of statistical estimators. By applying direct standardization, the MQI avoid the shortcomings of indirect standardization and provide easily interpretable results. In addition to defining MQI, we used Monte Carlo simulation to assess their performance relative to established approaches to quality measurement in healthcare.

\hypertarget{methods}{%
\section{Methods}\label{methods}}

\hypertarget{statmod}{%
\subsection{Statistical model}\label{statmod}}

We consider patients \(i=1,…,n\) living in patient regions \(r=1,…,R\). Each patient was treated because of a specific medical condition (e.g., stroke) at a hospital \(h=1,…H\) located in hospital region \(s=1,…,S\). We refer to hospitals for convenience only. Depending on the context, \(h\) may denote other types of healthcare providers.

Note that patient regions \(r\) and hospital regions \(s\) do not have to be identical. In our notation, subscripts indicate patient characteristics and superscripts indicate hospital characteristics. Thus, we use \(H_r\) to denote the number of hospitals that treated patients living in region \(r\) and \(H^s\) to denote the number of hospitals in hospital region \(s\). In a similar way, we write \(n_r^h\) to denote the number of patients living in region \(r\) who were treated at hospital \(h\) and \(n_r^s\) to denote the number of patients living in region \(r\) who were treated at hospitals located in region \(s\).

For each patient \(i\) living in region \(r\) who was treated at hospital \(h\), we consider an outcome \(y_{ri}^h | \mu_{ri}^h \overset{\text{ind.}}{\sim} G(\mu_{ri}^h)\), which is distributed according to \(G(\cdot)\) with expected value \(\mu_{ri}^h\). Examples of possible outcomes in real world applications include in-hospital mortality, complications after surgery, or hospital readmission. We model the expected value of \(y_{ri}^h\) as

\begin{align}
\mu_{ri}^h = E[y_{ri}^h|\boldsymbol{x}_{ri}^h,\boldsymbol{\beta},\theta^h,\eta_r] = F^{-1}({\boldsymbol{x}'}_{ri}^h \boldsymbol{\beta} + \theta^h + \eta_r) ,
\label{eq:prhi}
\end{align}

where \(\boldsymbol{x}_{ri}^h\) represents patient-specific risk factors with coefficient vector \(\boldsymbol{\beta}\), which are linked to patient outcomes via the inverse link function \(F^{-1}(\cdot)\). In addition to patient characteristics, \eqref{eq:prhi} includes a hospital-specific effect \(\theta^h\) and a patient-region-specific effect \(\eta_r\). \(\theta^h\) represents the influence of hospital characteristics such as case volume or the availability of specialized treatments on patient outcomes. Thus, \(\theta^h\) captures outcome-related hospital quality of care. \(\eta_r\) represents the influence of regional characteristics on patient outcomes. Those characteristics may, for instance, include accessibility of healthcare services.

Our model exploits available information on hospital characteristics by including them as predictors. Given hospital characteristics \(\boldsymbol{z}^h\), we model the hospital-specific effect as

\begin{equation}
 \theta^h = {\boldsymbol{z}'}^h \boldsymbol{\gamma} + u^h ,
 \label{eq:theta}
\end{equation}

where \(\boldsymbol{\gamma}\) is a vector of coefficients and \(u^h \overset{i.i.d.}{\sim} N(0,\sigma_u^2)\) is a random effect that represents unobserved heterogeneity between hospitals.

In a similar fashion, we model the patient-region effect as

\begin{equation}
 \eta_{r} = \boldsymbol{w}'_{r} \boldsymbol{\delta} + v_{r} + \phi_r ,
 \label{eq:eta}
\end{equation}

where \(\boldsymbol{w}_{r}\) are regional characteristics with coefficients \(\boldsymbol{\delta}\) and \(v_r \overset{i.i.d.}{\sim} N(0,\sigma_v^2)\) is a random effect capturing unobserved heterogeneity between patient regions. As an additional term, \eqref{eq:eta} includes an intrinsic conditional auto-regressive (ICAR) component \(\phi_r\), which is a spatially structured random effect that captures correlation between patient outcomes in neighboring patient regions. Such correlation may, for instance, arise when patients living in different regions use healthcare services from the same providers. Another cause of such correlation may be a high level of air pollution in one region that also affects patient outcomes in neighboring regions.

Expressions \eqref{eq:prhi}-\eqref{eq:eta} represent a multilevel model.\textsuperscript{23} This model may be written in a more simplified way by inserting \eqref{eq:theta} and \eqref{eq:eta} into \eqref{eq:prhi}:

\begin{equation}
\mu_{ri}^h = F^{-1}({\boldsymbol{x}'}_{ri}^h \boldsymbol{\beta} + [{\boldsymbol{z}'}^h \boldsymbol{\gamma} + u^h] + [\boldsymbol{w}'_{r} \boldsymbol{\delta} + v_{r} + \phi_r] ) .
 \label{eq:prhi2}
\end{equation}

The expected outcome of a patient is thus modeled as a function of patient, hospital, and regional characteristics. The coefficients \(\boldsymbol{\gamma}\) and \(\boldsymbol{\delta}\) may be used to directly assess relationships between specific characteristics of hospitals/regions and patient outcomes. For the ultimate purpose of comparison of hospitals and regions and for public reporting, we use the indicators defined in the following.

\hypertarget{definition-of-mqi}{%
\subsection{Definition of MQI}\label{definition-of-mqi}}

Based on the model described above, we propose three indicators: one at the hospital level and two at the regional level. All indicators answer specific questions, which are formulated below. In contrast to approaches relying on indirect standardization by using measures such as the SMR, the proposed indicators use direct standardization by averaging outcomes across the whole patient population.\textsuperscript{15}

\uline{Standardized hospital outcome rate (SHOR)}:

\textbf{Question}: Consider a hospital indexed by \(h\). What is the expected outcome rate (e.g., the expected mortality, complication, or readmission rate) if all patients had been treated at a hospital identical to the considered hospital?

\textbf{Answer}: The SHOR of hospital \(h\) is defined as

\begin{equation}
\text{SHOR}^h = \frac{1}{n} \sum_{r^*=1}^R \sum_{h^*=1}^{H_{r^*}} \sum_{i=1}^{n_{r^*}^{h^*}} F^{-1}({\boldsymbol{x}'}_{r^*i}^{h^*} \boldsymbol{\beta} + [{\boldsymbol{z}'}^h \boldsymbol{\gamma} + u^h] + [\boldsymbol{w}'_{r^*} \boldsymbol{\delta} + v_{r^*} + \phi_{r^*}] ) .
 \label{eq:SHOR}
\end{equation}

Note that the hospital effect \({\boldsymbol{z}'}^h \boldsymbol{\gamma} + u^h\) in \eqref{eq:SHOR} is specific to the considered hospital \(h\) whereas characteristics of patients and regions vary according to the original data. This is equivalent to the hypothetical situation in which the outcome-relevant characteristics of the considered hospital are assigned to all hospitals while keeping case mix and regional effects unchanged. Alternatively, one may think of \eqref{eq:SHOR} as the hypothetical result of opening one new hospital with the same characteristics as the considered hospital in each region. The SHOR then is the average outcome rate if all patients in all regions were allocated to these hypothetical new hospitals.

\uline{Regional standardized hospital outcome rate (RSHOR)}:

\textbf{Question}: Consider a hospital region indexed by \(s\). What is the expected outcome rate (e.g., the expected mortality, complication, or readmission rate) if all patients had been treated at hospitals in the considered region?

\textbf{Answer}: The RSHOR of region \(s\) is defined as

\begin{equation}
\text{RSHOR}^s = \sum_{h=1}^{H^s} \frac{n^{h}}{n^s} \cdot \text{SHOR}^h,
\label{eq:rSHOR}
\end{equation}

where \(n^{h}/n^s\) is the share of patients treated in region \(s\) that were treated at hospital \(h\).

The RSHOR is the patient-share weighted mean of the SHORs of the hospitals located in the considered region. This corresponds to a hypothetical situation in which all patients are allocated to the hospitals of that region according to these hospitals' actual shares of patients treated in that region. Thus, the RSHOR makes statements about the average quality of care in regional hospitals.

\uline{Regional standardized patient outcome rate (RSPOR)}:

\textbf{Question}: Consider a patient region indexed by \(r\). What is the expected outcome rate (e.g., the expected mortality, complication, or readmission rate) if all patients had been living in that region?

Answer: The RSPOR of region \(r\) is defined as

\begin{equation}
\text{RSPOR}_r = \sum_{h=1}^{H_r} \frac{n_r^{h}}{n_r} \cdot \bar{p}_r^h,
\label{eq:rpqi}
\end{equation}

where \(n_r^h/n_r\) is the share of patients living in region \(r\) who were treated at hospital \(h\) and

\begin{equation}
\bar{p}_r^h = \frac{1}{n} \sum_{r^*=1}^R \sum_{h^*=1}^{H_{r^*}} \sum_{i=1}^{n_{r^*}^{h^*}} F^{-1}({\boldsymbol{x}'}_{r^*i}^{h^*} \boldsymbol{\beta} + [{\boldsymbol{z}'}^h \boldsymbol{\gamma} + u^h] + [\boldsymbol{w}'_{r} \boldsymbol{\delta} + v_{r} + \phi_{r}] )
 \label{eq:prhe}
\end{equation}

is the hypothetical average outcome rate (e.g., the expected mortality, complication, or readmission rate) of patients living in region \(r\) who were treated at hospital \(h\).

The RSPOR is a patient-share weighted arithmetic mean of the expected outcome rates of the hospitals that treated patients living in the considered region. This corresponds to a hypothetical situation in which all patients in the sample are living in region \(r\) and are allocated to hospitals according to these hospitals' actual shares of treated patients from the considered region. Contrary to \eqref{eq:SHOR}, \eqref{eq:prhe} includes both the effect of the considered hospital \({\boldsymbol{z}'}^h \boldsymbol{\gamma} + u^h\) and effect of the considered patient region \(\boldsymbol{w}'_{r} \boldsymbol{\delta} + v_{r} + \phi_{r}\). This ensures that regional characteristics, which are related to patient outcomes, are taken into account. Patient characteristics in \eqref{eq:prhe} vary according to the original data, which ensures that the same case mix is underlying the comparison of regions.

The RSHOR and the RSPOR represent different, complementary perspectives on regional healthcare. By focusing on hospital performance in a specific region, the RSHOR represents a supply-oriented perspective. This perspective may be useful to identify potentials for improvement in regional inpatient care. On the contrary, the RSPOR represents a demand-oriented perspective as it describes the standardized outcomes of patients living in a specific region. This perspective may be relevant to identify those regional populations with the highest need for improvements in healthcare and living conditions.

\hypertarget{data-generation-process}{%
\subsection{Data generation process}\label{data-generation-process}}

We used Monte Carlo simulation to assess the statistical properties of MQI methodology. We also compared MQI methodology to conventional approaches to the design of quality indicators. In contrast to real-world applications, Monte Carlo simulation offers the advantage that the data generation process (DGP) is fully specified by the researcher. This implies that the true parameter values, including those related to quality of care, are known. These true parameter values, in turn, provide the basis to assess the performance of different estimation approaches.

Our DGP considered \(R\) regions, which were characterized by the same number of hospitals \(\bar{H}\). Hence, the total number of simulated hospitals was \(H=\bar{H} \cdot R\). For simplicity, we assumed that all patients living in a specific region \(r\) were treated at a hospital located in that region. We also did not consider inter-regional dependencies as captured by the ICAR component \(\phi_r\). Given these simplifications, we defined the region-specific effect as

\begin{equation}
 \eta_{r} = \delta w_r + v_{r} ,
 \label{eq:etaspec}
\end{equation}

where \(w_r\overset{\text{i.i.d.}}{\sim}\text{Ber}(0.5)\) is a binary, region-specific, observed variable (e.g., urban/rural area) with expected value \(E[w_r]=p_w=0.5\) and variance \(V[w_r]=\sigma_w^2=0.25\) and \(v_r\overset{\text{i.i.d.}}{\sim}N(0,\sigma_v^2)\) is an unobserved term.

To determine the value of \(\delta>0\), we fixed the share of the variance of \(\eta_r\) induced by \(w_r\) to \(\xi_{w_r}^{\eta_r}=V[\delta w_r]/V[\eta_r]\). Accordingly, \(\delta\) was derived as

\begin{equation}
\delta = \frac{\sigma_v}{\sigma_w} \sqrt{\frac{\xi_{w_r}^{\eta_r}}{1-\xi_{w_r}^{\eta_r}}}  .
 \label{eq:delta}
\end{equation}

Thus, \eqref{eq:delta} was used to determine \(\delta\) by setting the share \(\xi_{w_r} ^{\eta_r}\)of variance caused by observed influence factors at the regional level. By keeping region-level variation constant at \(\sigma_{\eta}^2 = V[\eta_r]\), we also derived \(\sigma_v^2 = (1-\xi_{w_r}^{\eta_r}) \sigma_{\eta}^2\).

At the hospital level, the DGP considered hospital case volume \(n^h\) as observed influence factor, i.e.,

\begin{equation}
 \theta^h = \gamma n^h + u^h ,
 \label{eq:thetaspec}
\end{equation}

where \(u^h\overset{\text{i.i.d.}}{\sim}N(0,\sigma_u^2)\) captures unobserved heterogeneity at the hospital level. The DGP allowed hospital volume to differ systematically between regions. This was achieved by generating hospital volume according to a conditional discrete uniform distribution

\begin{equation}
n^h | \lambda_r \overset{\text{ind.}}{\sim} \text{Uni}(1,\lambda_r) ,
\end{equation}

where the maximum hospital volume \(\lambda_r\) was given by

\begin{equation}
\lambda_r = 
\begin{cases} 
\lambda_{r0} = 2\bar{n}-1-\Delta n / 2 : w_r = 0 \\
\lambda_{r1} = 2\bar{n}-1+\Delta n / 2 : w_r = 1
\end{cases} .
\end{equation}

Here, \(\Delta n\) (\(|\Delta n| \leq 4(\bar{n}-1)\)) is an even number governing the difference between maximum hospital volumes in regions with \(w_r=0\) and regions with \(w_r=1\). Thus, a higher (absolute) value of \(\Delta n\) induces higher heterogeneity of regions with respect to hospital volume.

Given the generating equations and assumptions outlined above, the expected hospital volume is \(E[n^h] = \bar{n}\) and the expected total number of simulated patients amounts to \(E[n]=\bar{n} \cdot H\). Thus, expected sample size is independent from \(\Delta n\).

Analogous to \(\delta\), we determined the value of \(\gamma<0\) by setting the variance of \(\theta^h\) induced by the observed hospital-specific influence factors to \(\xi_{n^h}^{\theta^h}=V[\gamma n^h]/V[\theta^h]\). Solving for \(\gamma\) yields

\begin{equation}
\gamma = - \frac{\sigma_u}{\sigma_{n^h}} \sqrt{\frac{\xi_{n^h}^{\theta^h}}{1-\xi_{n^h}^{\theta^h}}} ,
\end{equation}

where

\begin{equation}
\sigma_{n^h}^2 = V[n^h] = \frac{1}{24} ( \lambda_{r1}^2 + \lambda_{r0}^2 - 2) + \frac{1}{16} (\lambda_{r1} - \lambda_{r0})^2 .
\end{equation}

By imposing that \(\gamma<0\), our DGP induced better patient outcomes in hospitals with higher case volumes and, thus, assumed the presence of volume-outcome relationships. Given the above expressions, we kept the variance of the hospital effect constant at \(\sigma_{\theta}^2 = V[\theta^h]\) and used the relation \(\sigma_u^2 = (1-\xi_{n^h}^{\theta^h}) \sigma_{\theta}^2\) to determine the variance of the term capturing unobserved hospital-level variation.

At the patient level, we considered a continuous risk factor \(x_{ri}^h | \mu_x^h,\sigma_x^2 \overset{\text{ind.}}{\sim} N(\mu_x^h, \sigma_x^2)\), where \(\mu_x^h\) is a hospital-specific expected value and \(\sigma_x^2\) represents within-hospital variance of patient risk. By letting the expected value \(\mu_x^h\) vary between hospitals, our DGP allowed for systematic differences between hospitals in terms of case mix. In addition, we induced possible correlation \(\rho=\text{Corr}[\mu_x^h,n^h]\) between the expected value of the patient-specific risk factor and hospital case volume by specifying

\begin{equation}
\mu_x^h = \chi n^h + \varepsilon^h ,
 \label{eq:muxhspec}
\end{equation}

where \(\varepsilon^h \overset{\text{i.i.d.}}{\sim} N(0, \sigma_\varepsilon^2)\) represents hospital-specific random variation. Given a specified value of \(\rho\), we set the value of \(\chi\) by using

\begin{equation}
\chi = \text{sign}(\rho) \frac{\sigma_\varepsilon}{\sigma_{n^h}} \sqrt{\frac{\rho^2}{1-\rho^2}} .
\end{equation}

The patient-specific risk factor, the hospital-specific effect, and the region-specific effect were related to a binary patient outcome \(y_{ri}^h | p_{yri}^h \overset{\text{ind.}}{\sim}\text{Ber}(p_{yri}^h)\) (e.g., in-hospital-mortality: no/yes), where

\begin{equation}
p_{yri}^h = \text{logit}^{-1}(\alpha + x_{ri}^h + \theta^h + \eta_r) 
 \label{eq:pyspec}
\end{equation}

is the outcome probability of patient \(i\) living in region \(r\) who was treated at hospital \(h\). While the coefficient of \(x_{ri}^h\) in \eqref{eq:pyspec} was set to unity, the importance of patient risk relative to hospital effects was governed by the variance parameter \(\sigma_\varepsilon^2\). We defined \(\xi_{\theta^h}^{\mu_x^h}=V[\mu_x^h]/V[\theta^h]\) as the variance of hospital-specific case mix differences relative to the variance of the hospital effect. Given this definition, we set

\begin{equation}
\sigma_\varepsilon^2 = \xi_{\theta^h}^{\mu_x^h} \cdot (1-\rho^2) \cdot \sigma_{\theta}^2
\end{equation}

to keep this variance ratio constant.

Another crucial parameter in our DGP is the average outcome probability \(\bar{p}_y=E[p_{yri}^h]\). To keep this probability constant across different scenarios, we determined the constant term \(\alpha\) by applying first-order Taylor series expansion of \eqref{eq:pyspec} around the point \(\text{logit}(\bar{p}_y)\):

\begin{equation}
\bar{p}_y \approx \bar{p}_y + \bar{p}_y \cdot (1-\bar{p}_y) \cdot \left( \alpha + E[x_{ri}^h] + E[\theta^h] + E[\eta_r] - \text{logit}(\bar{p}_y) \right) .
\end{equation}

Solving for \(\alpha\) yields

\begin{align}
\alpha &\approx \text{logit}(\bar{p}_y) - E[x_{ri}^h] - E[\theta^h] - E[\eta_r] \\
&= \text{logit}(\bar{p}_y) - (\chi + \gamma) E[n^h] - \zeta \delta
 \label{eq:alpha}
\end{align}

where

\begin{equation}
E[n^h] = \frac{1}{3} [\zeta \cdot (2\lambda_{r1}+1) + (1-\zeta) \cdot (2\lambda_{r0}+1)]    
\end{equation}

and

\begin{equation}
\zeta = \frac{\lambda_{r1}+1}{\lambda_{r0}+\lambda_{r1}+2}
\end{equation}

When using \eqref{eq:alpha}, it is noteworthy that the expectation of \(n^h\) at the patient level differs from the expectation of \(n^h\) at the hospital level. The reason for this deviation is that a hospital with volume \(n^h\) is represented not once but with \(n^h\) patients in each final simulated dataset. Accordingly, \(n^h|\lambda_r\) is uniformly distributed at the hospital level but not at the patient level. The same argument applies to the expected value of \(\eta_r\).

\hypertarget{scenarios}{%
\subsection{Scenarios}\label{scenarios}}

\hypertarget{baseline-scenario}{%
\subsubsection{Baseline scenario}\label{baseline-scenario}}

For simulation, we defined a baseline scenario, which was characterized by the parameter values shown in Table \ref{tab:base}. We considered \(R=20\) regions, each with \(\bar{H}=10\) hospitals. The average case volume was set to \(\bar{n}=10\), resulting in an expected sample size of \(E[n]=2,000\). The average outcome probability was set to 30\% (\(\bar{p}_y=0.3\)). The baseline scenario assumed no systematic differences between regions in terms of hospital size (\(\Delta n =0\)) and no correlation between case mix and hospital volume (\(\rho=0\)). The shares of the total variance induced by observed variables were set to 50\% at both the hospital level (\(\xi_{n^h}^{\theta^h}=0.5\)) and the regional level (\(\xi_{w_r}^{\eta_r}=0.5\)). Case mix differences between hospitals and the hospital effect were assumed to have equal variance (\(\xi_{\theta^h}^{\mu_x^h}=1\)). We chose equal variance parameter values at the regional and the hospital level (\(\sigma_\eta = \sigma_{\theta}= 0.5\)) and a smaller within-hospital case mix variance (\(\sigma_x=0.2\)).

\begin{table}[b] \centering \caption{Baseline parameter values \label{tab:base}}
\begin{tabular}{rrrrrrrrrrrr} \toprule
$R$ & $\bar{H}$ & $\bar{n}$ & $\bar{p}_y$ & $\Delta n$ & $\rho$ & $\xi_{w_r}^{\eta_r}$ & $\xi_{n^h}^{\theta^h}$ & $\xi_{\theta^h}^{\mu_x^h}$ & $\sigma_\eta$ & $\sigma_{\theta}$ & $\sigma_x$ \\ \hline
20 & 10 & 10 & 0.3 & 0 & 0 & 0.5 & 0.5 & 1 & 0.5 & 0.5 & 0.2 \\ \bottomrule
\end{tabular}
\end{table}

\hypertarget{variations-of-parameter-values}{%
\subsubsection{Variations of parameter values}\label{variations-of-parameter-values}}

We investigated the influence of changes in specific parameter values on the performance of different quality indicators (Table \ref{tab:parvar}). The considered scenarios included different directions and strengths of correlation between case mix and case volume \(\rho\), different magnitudes of between-hospital case mix variation as governed by \(\xi_{\theta^h}^{\mu_x^h}\), different shares of variance at the hospital level induced by hospital case volume \(\xi_{n^h}^{\theta^h}\), different degrees of regional heterogeneity with respect to hospital case volume as determined by \(\Delta n\), different expected outcome probabilities \(\bar{p}_y\), and different standard deviations of the regional effect \(\sigma_\eta\). For each of the considered scenarios, we ran 1,000 simulations to obtain Monte Carlo estimates of the performance measures described below.

\begin{table}[tb]
  \centering  \caption{Parameter values used for simulation \label{tab:parvar}}
    \begin{tabular}{ll} \toprule
    Parameter & Values \\ \hline
    $\rho$ & $\{-0.8,-0.6,…,0.8\}$ \\
    $\xi_{\theta^h}^{\mu_x^h}$ & $\{0.2,0.5,0.75,1,1.25,1.5,2,5,7.5,10\}$ \\
    $\xi_{n^h}^{\theta^h}$ & $\{0.01,0.1,0.2,0.4,0.5,0.6,0.8,0.9,0.99\}$ \\
    $\Delta n$  & $\{-16,-10,-6,-2,0,2,6,10,16\}$ \\
    $\bar{p}_y$ & $\{0.03,0.05,0.1,0.15,0.2,0.3,0.4,0.5\}$ \\ 
    $\sigma_{\eta}$ & $\{0,0.1,0.25,0.5,0.75,1,2\}$ \\   \bottomrule
    \end{tabular}%
  \label{tab:addlabel}%
\end{table}%

\hypertarget{alternative-quality-indicators}{%
\subsection{Alternative quality indicators}\label{alternative-quality-indicators}}

\hypertarget{hospital-level-indicators}{%
\subsubsection{Hospital-level indicators}\label{hospital-level-indicators}}

In addition to the SHOR defined by \eqref{eq:SHOR}, we assessed alternative indicators of hospital performance. First, we calculated the raw mortality rates of all hospitals:

\begin{equation}
\bar{y}^h = \frac{1}{n^h} \sum_{i=1}^{n^h} y_{ri}^h .
\end{equation}

Since the raw mortality rate does not adjust for case mix, it is rarely used as a measure of hospital quality in practical applications. However, it is considered here as reference.

Second, we calculated the hospitals' SMRs as

\begin{equation}
\text{SMR}^h = \frac{\sum_{i=1}^{n^h} y_{ri}^h}{\sum_{i=1}^{n^h} \hat{p}_{yri}^{h}} ,
\end{equation}

where \(\hat{p}_{yri}^{h}\) is the patient's predicted outcome probability, which was derived from the logistic regression model

\begin{equation}
E[y_{ri}^h | x_{ri}^h] = \text{logit}^{-1}(b_0 + b_1 x_{ri}^h) .
\label{eq:lrsmr}
\end{equation}

Given the parameter estimates \(\hat{b}_0\) and \(\hat{b}_1\), \(\hat{p}_{yri}^{h}\) was calculated as

\begin{equation}
\hat{p}_{yri}^{h} = \text{logit}^{-1}(\hat{b}_0 + \hat{b}_1 x_{ri}^h) .
\label{eq:estpyrhi}
\end{equation}

Since this procedure of deriving hospital SMRs is frequently used to estimate hospital quality, it served as an important benchmark in our simulation.

We also considered the risk standardized mortality rate (RSMR) as used by the Centers for Medicare \& Medicaid Services (CMS).\textsuperscript{24} The RSMR relies on the random-intercept multilevel model

\begin{equation}
E[y_{ri}^h | x_{ri}^h,a^h] = \text{logit}^{-1}(a^h + b x_{ri}^h) ,
\label{eq:lrrsmr}
\end{equation}

where \(a^h \overset{\text{i.i.d.}}{\sim} N(\bar{a},\sigma_a^2)\) is a normally distributed random effect. Given the estimates \(\hat{a}^h\), \(\hat{\bar{a}}\), and \(\hat{b}\), the RSMR was derived as

\begin{equation}
\text{RSMR}^h = \frac{\sum_{i=1}^{n^h} \text{logit}^{-1}(\hat{a}^h + \hat{b} x_{ri}^h)}{\sum_{i=1}^{n^h} \text{logit}^{-1}(\hat{\bar{a}} + \hat{b} x_{ri}^h)} .
\end{equation}

Finally, we investigated the effect of omitting the regional effect from the model underlying the SHOR by specifying the expected outcome of a patient as

\begin{equation}
\mu_{ri}^h = F^{-1}({\boldsymbol{x}'}_{ri}^h \boldsymbol{\beta} + [{\boldsymbol{z}'}^h \boldsymbol{\gamma} + u^h]) .
\end{equation}

The performance of the SHOR without regional effects derived from this model was used to assess the relevance of modeling the regional level for the adequacy of hospital-level estimates.

\hypertarget{area-level-indicators}{%
\subsubsection{Area-level indicators}\label{area-level-indicators}}

At the regional level, we compared the performance of the RSPOR \eqref{eq:rpqi} with the regional SMR, which was defined as

\begin{equation}
\text{SMR}_r = \frac{\sum_{i=1}^{n_r} y_{ri}^h}{\sum_{i=1}^{n_r} \hat{p}_{yri}^{h}} ,
\end{equation}

where the estimate \(\hat{p}_{yri}^{h}\) was given by \eqref{eq:estpyrhi}.

\hypertarget{performance-assessment}{%
\subsection{Performance assessment}\label{performance-assessment}}

We used three performance measures to assess the adequacy of the different approaches to estimate hospital quality:

\begin{enumerate}
\def\labelenumi{\arabic{enumi}.}
\item
  Spearman's rank correlation coefficient between the true hospital effect \(\theta^h\) and the considered hospital quality indicator,
\item
  the share of the 10\% best hospitals identified by the hospital quality indicator,
\item
  the share of the 10\% worst hospitals identified by the hospital quality indicator.
\end{enumerate}

For all of these performance measures, higher values indicate better correspondence of estimation results with true hospital quality. While the rank correlation coefficient evaluates the adequacy of the full hospital ranking implied by a specific indicator, the shares of the 10\% worst and 10\% best hospitals focus on the tails of the hospital quality distribution.

Due to the small number of simulated regions, we only used Spearman's rank correlation with \(\eta_r\) to asses the performance of area-level indicators.

\hypertarget{results}{%
\section{Results}\label{results}}

\hypertarget{rank-correlation-with-true-hospital-effects}{%
\subsection{Rank correlation with true hospital effects}\label{rank-correlation-with-true-hospital-effects}}

\begin{figure}
\includegraphics[width = .95\textwidth]{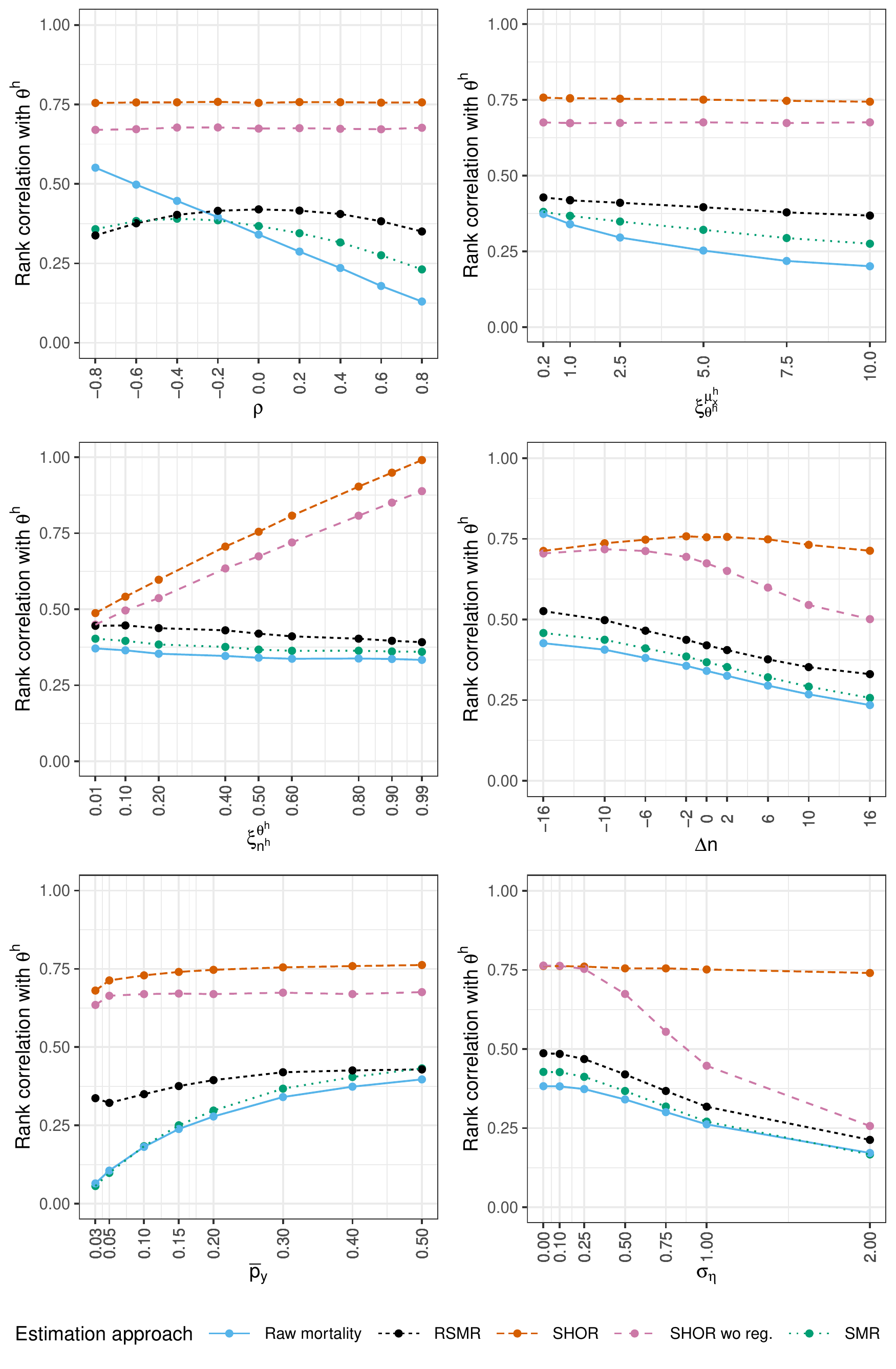}
\caption{Rank correlation between the true hospital effect $\theta^h$ and quality indicators under different scenarios \label{fig:rc}}
\end{figure}

In all considered scenarios, the SHOR outperformed the raw mortality rate, the SMR, and the RSMR in terms of rank correlation with the true hospital effect \(\theta^h\) (Figure \ref{fig:rc}). The SHOR based on the model including regional effects consistently showed better or equal performance compared with the SHOR based on the model without regional effects. In most scenarios, the RSMR yielded a more adequate hospital ranking than the raw mortality rate and the SMR.

The SHOR was robust against correlation between hospital case mix and case volume \(\rho\) whereas the SMR, the RSMR, and particularly the raw mortality rate were affected by changes in this parameter (upper left panel of Figure \ref{fig:rc}). Since case volume entered our DGP as a component of hospital quality, the finding that the SMR is sensitive to such correlation is consistent with those of previous simulations highlighting distortions due to correlation between hospital quality and patient-specific risk factors.\textsuperscript{6} The robustness of the SHOR results from the fact that both the patient-specific risk factor \(x_{ri}^h\) and case volume \(n^h\) enter the multilevel model \eqref{eq:prhi} as regressors. Hence, correlation between these variables is captured when estimating the model parameters.

The SHOR also proved to be robust against larger case mix differences between hospitals as induced by higher values of \(\xi_{\theta^h}^{\mu_x^h}\) (upper right panel of Figure \ref{fig:rc}). In contrast, the performance of the raw mortality rate, the SMR, and the RSMR declined in between-hospital patient heterogeneity. The latter result is notable, since it demonstrates that case mix adjustment based on the SMR and the RSMR does not fully account for differences between hospitals although patient risk \(x_{ri}^h\) has been fully observed and was included in the logistic regression models \eqref{eq:lrsmr} and \eqref{eq:lrrsmr}.

A higher share of the variance of the hospital effect induced by hospital case volume \(\xi_{n^h}^{\theta^h}\) was related to better performance of the SHOR (middle left panel of Figure \ref{fig:rc}). Intuitively, this finding reflects that estimation of hospital quality becomes more accurate when more relevant hospital characteristics related to patient outcomes are included in the model \eqref{eq:prhi} underlying the SHOR. Notably, the SHOR based on the model with regional effects also performed better than the other measures when case volume was assumed to account for a very low share of the variance of \(\theta^h\).

Higher regional heterogeneity in terms of hospital size affected the performance of all estimation approaches (middle right panel of Figure \ref{fig:rc}). However, the SHOR based on the model including regional effects was most robust to such heterogeneity. Both positive and negative values of \(\Delta n\) were related to slightly worse performance of the SHOR including regional effects. This worse performance is caused by the higher variance of the estimator of \(\gamma\) that results from correlation between the region-specific variable \(w_r\) and hospital case volume \(n^h\) as induced by \(\Delta n \neq 0\). The SHOR without regional effects was strongly affected by changes in \(\Delta n\), which reflects that correlation between the hospital effect and omitted regional effect induces bias in the estimator of \(\gamma\). The raw mortality rate, the SMR, and the RSMR showed better performance for negative and worse performance for positive values of \(\Delta n\).

Since estimators of model coefficients generally show high variance in the case of rare outcomes, a higher average outcome probability \(\bar{p}_y\) induced better performance of all estimation approaches (lower left panel of Figure \ref{fig:rc}). Again, the SHOR was remarkably more stable than its competitors. While the rank correlation with the hospital effect \(\theta^h\) approached zero for the raw mortality rate and the SMR when the average outcome probability was small, the SHOR still yielded a reasonably well ranking of hospitals. Hence, these results indicate that adequate estimation of hospital quality based on the SHOR may be possible even for rare outcomes.

Variations of the standard deviation of the regional effect \(\sigma_\eta\) particularly affected the performance of the SHOR based on the model without regional effects (lower right panel of Figure \ref{fig:rc}). While the adequacy of hospital rankings based on the SHOR with and without regional effects was similar when the variance of the regional effect was low, the performance of the SHOR without regional effects declined dramatically when \(\sigma_\eta\) was increased. This result reflects the higher variance of the estimator of \(\gamma\) caused by unobserved heterogeneity. In contrast, the performance of the SHOR with regional effects was hardly affected by changes in \(\sigma_\eta\) because the underlying model captured the higher variance at the regional level.

\begin{figure}[tb]
\includegraphics[width = .93\textwidth]{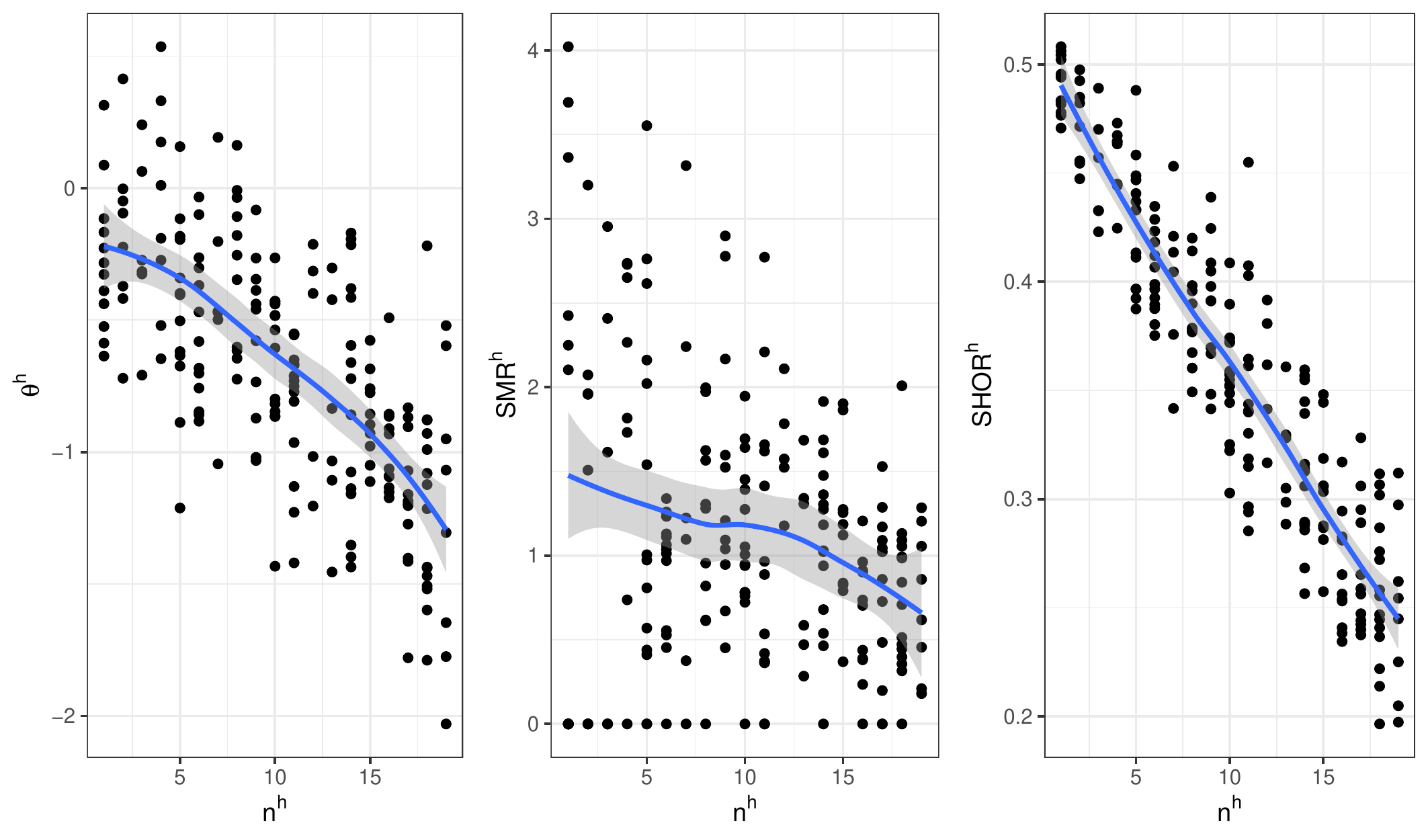}
\caption{Results of a typical simulation run of the baseline scenario \label{fig:examplebaseline}}
\end{figure}

Further insights into differences between the SHOR and the SMR may be gained by considering data from a typical simulation run of the baseline scenario (Figure \ref{fig:examplebaseline}). The left panel of the figure illustrates the relationship between the true hospital effect \(\theta^h\) and case volume \(n^h\) in this simulated dataset. The volume-outcome relationship induced by \eqref{eq:thetaspec} was revealed by the LOESS estimate for these simulated data. As shown in the middle panel of the figure, this volume-outcome relationship was less pronounced when considering the SMR. In addition, SMR estimates showed a relatively high variance, particularly for small-volume hospitals. In contrast, the SHOR estimates clearly revealed the volume-outcome relationship as shown in the right panel of the figure. The variability of the SHOR estimates along the LOESS estimate was relatively small and reached its minimum for low-volume hospitals. This is a consequence of the random effects estimator underlying the SHOR, which shrinks estimates towards the (conditional) mean.\textsuperscript{15,25}

\hypertarget{identification-of-best-and-worst-hospitals}{%
\subsection{Identification of best and worst hospitals}\label{identification-of-best-and-worst-hospitals}}

\begin{figure}
\includegraphics[width = .95\textwidth]{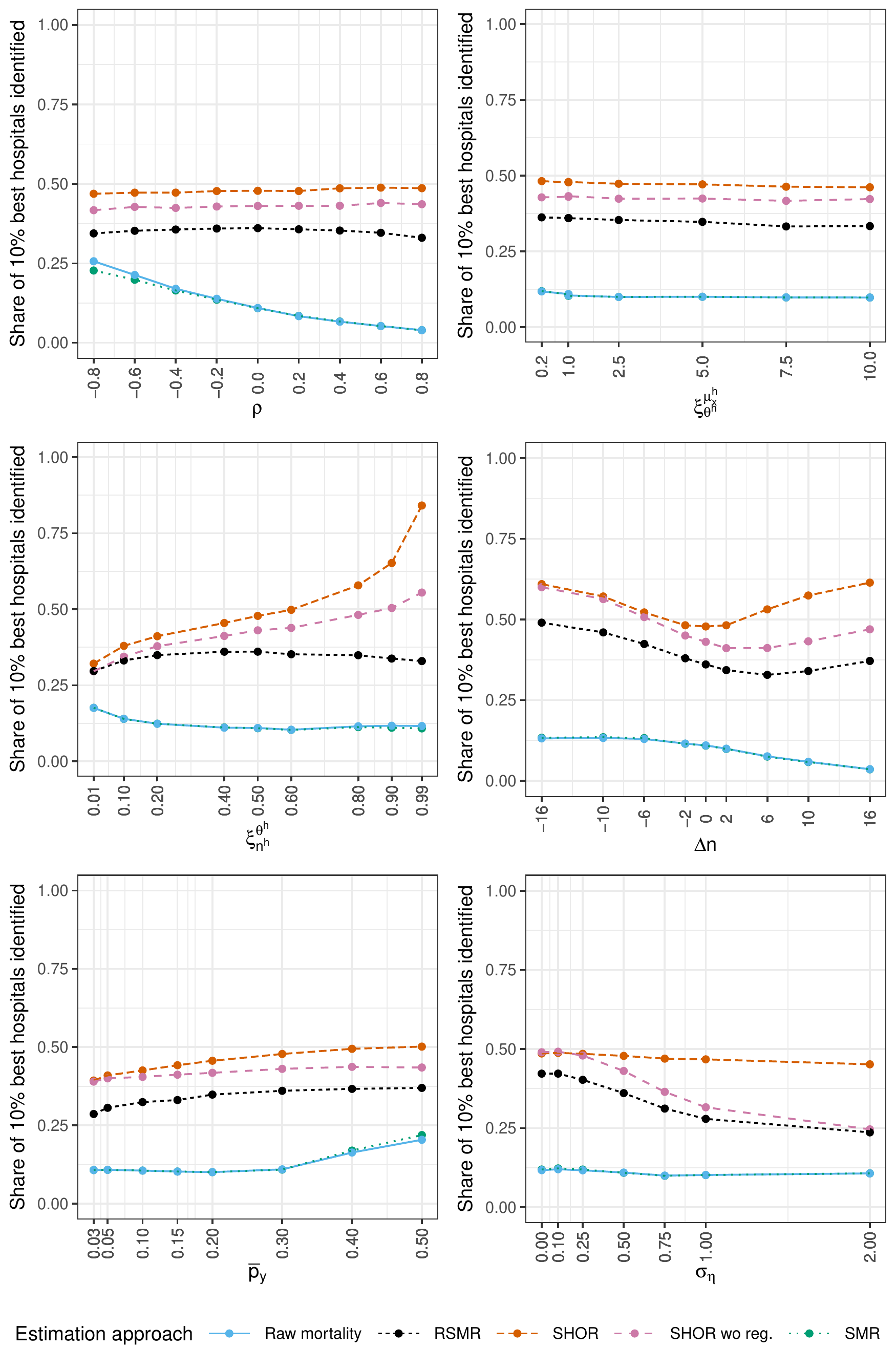}
\caption{Share of the 10\% best hospitals identified und different scenarios \label{fig:best}}
\end{figure}

\begin{figure}
\includegraphics[width = .95\textwidth]{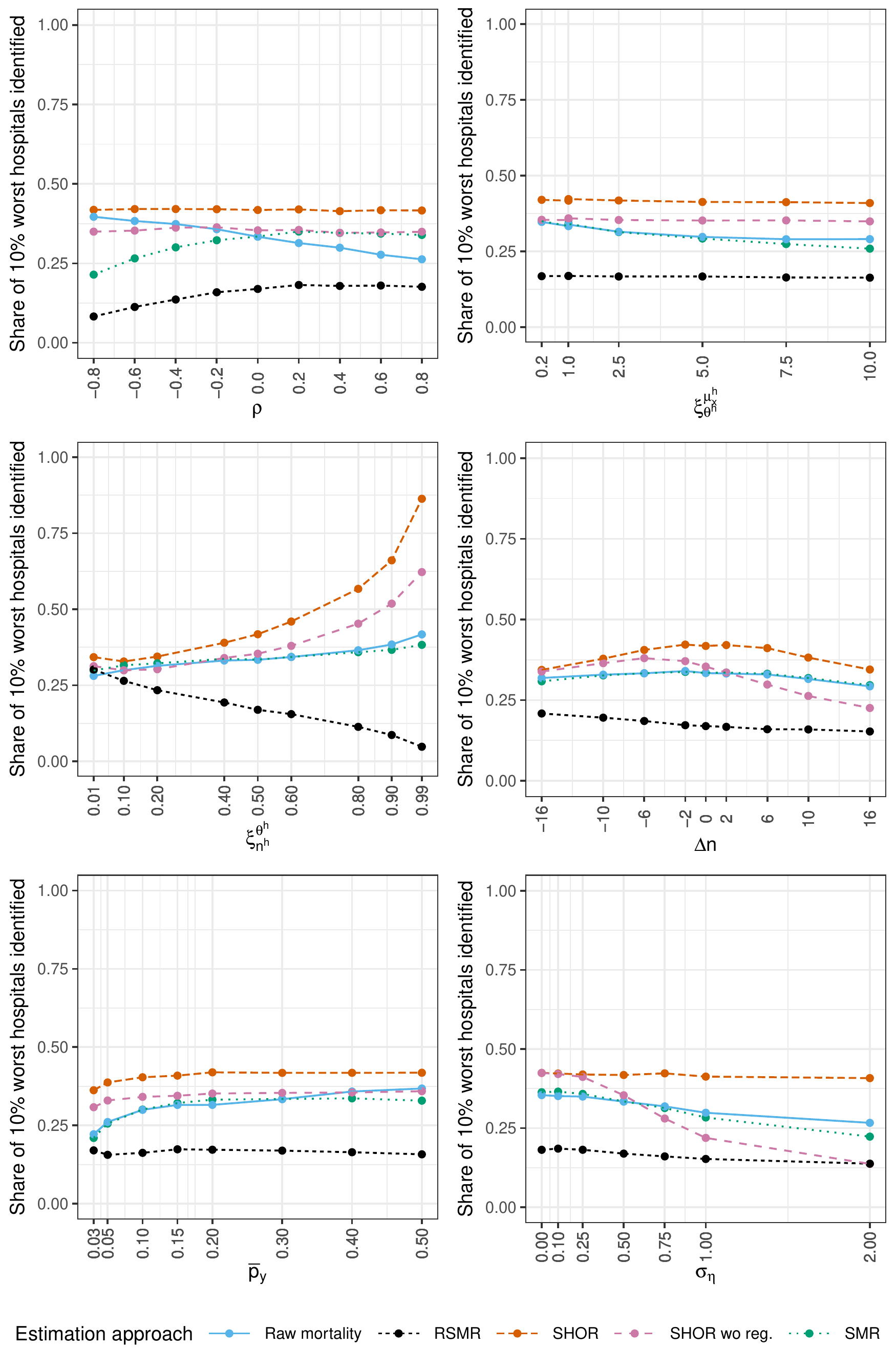}
\caption{Share of the 10\% worst hospitals identified und different scenarios \label{fig:worst}}
\end{figure}

The finding that the SHOR gives more adequate estimates of hospital quality than the raw mortality rate, the SMR, and the RSMR was also confirmed when considering the shares of the 10\% best hospitals (Figure \ref{fig:best}) and the 10\% worst hospitals (Figure \ref{fig:worst}) identified by these estimation approaches. Interestingly, there was little difference between the performances of the raw mortality rate and the SMR under most scenarios. In contrast, the SHOR performed obviously better in almost all cases. In our scenarios, those differences in performance were more pronounced when aiming to identify the best hospitals, particularly due to a relatively poor performance of the raw mortality rate and the SMR. However, this finding is likely driven by the link between hospital quality and case volume in the DGP and may not generalize to other settings.

\hypertarget{rank-correlation-with-true-region-effects}{%
\subsection{Rank correlation with true region effects}\label{rank-correlation-with-true-region-effects}}

\begin{figure}
\includegraphics[width = .95\textwidth]{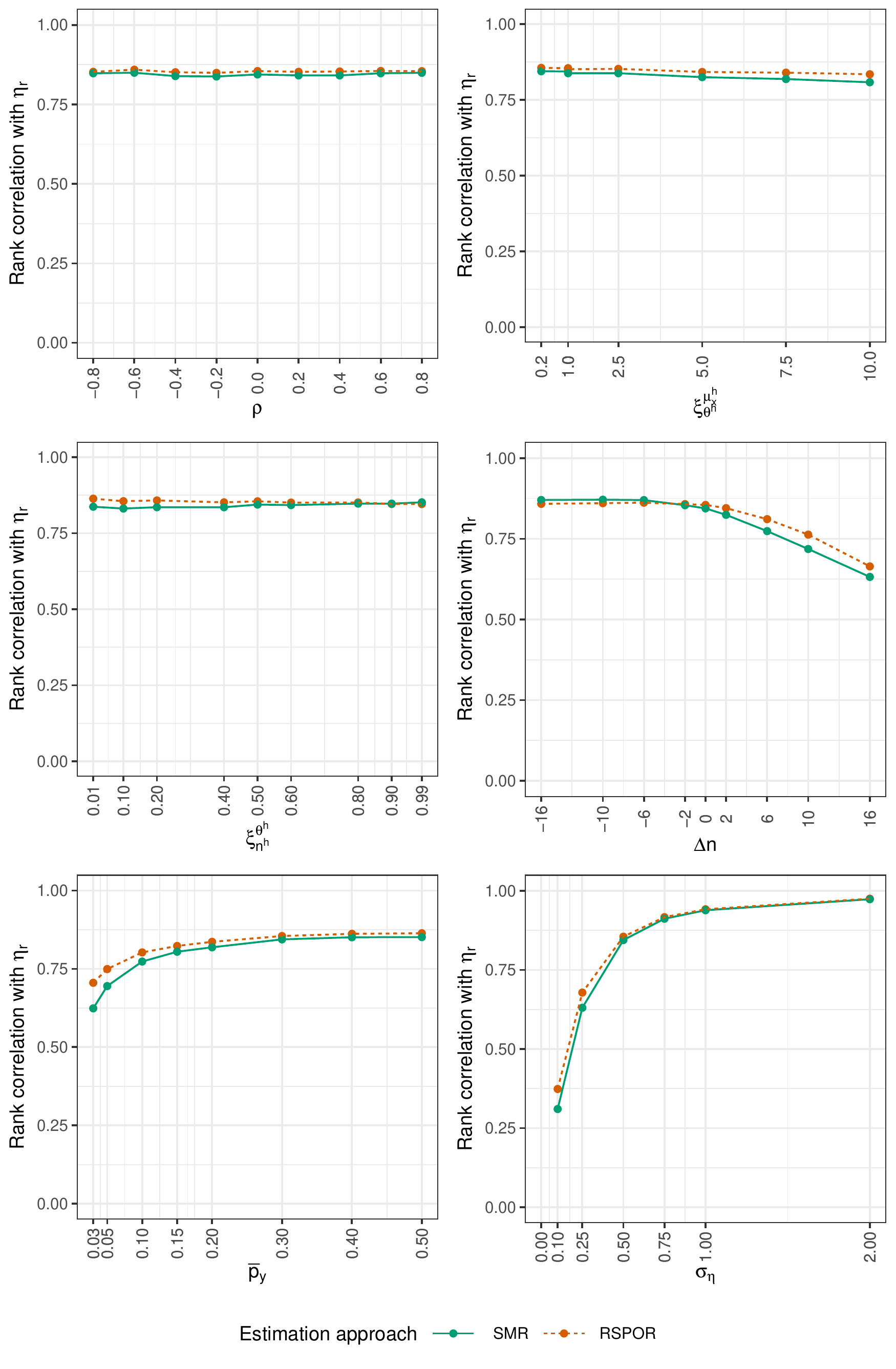}
\caption{Rank correlation between the true region effect $\eta_r$ and quality indicators under different scenarios \label{fig:rcreg}}
\end{figure}

At the regional level, our simulation results indicated only small performance differences between the RSPOR and the regional SMR (Figure \ref{fig:rcreg}). Both were robust against changes in hospital-level parameters (\(\rho\), \(\xi_{\theta^h}^{\mu_x^h}\),\(\xi_{n^h}^{\theta^h}\)). In contrast, changes in region-level parameters (\(\Delta n\), \(\sigma_\eta\)) and the average outcome rate \(\bar{p}_y\) affected the performance of both measures. The RSPOR generally showed slightly higher robustness compared with the regional SMR.

\hypertarget{discussion}{%
\section{Discussion}\label{discussion}}

This paper presented a methodological foundation of quality indicators for both healthcare providers and geographical regions. These Multilevel Quality Indicators (MQI) are derived from a single statistical model and facilitate comprehensive and consistent estimation of healthcare quality at different levels of aggregation.

Monte Carlo simulations showed that the proposed standardized hospital outcome rate (SHOR) provides more adequate and robust estimates of healthcare provider performance than established approaches based on the SMR and the RSMR. This result holds true under almost all considered scenarios. The main reason for the substantially better performance of the SHOR is the more efficient estimation of provider effects due to the inclusion of provider characteristics in the underlying statistical model. In addition, our results indicate that the inclusion of relevant regional characteristics in the statistical model also increases the adequacy of provider-level estimates.

At the regional level, our simulation results indicated that the regional standardized patient outcome rate (RSPOR) provides adequate estimates of regional differences and is slightly more robust than the regional SMR. However, due to the small number of simulated regions, our results for this area-level indicator should be considered as preliminary.

While our simulations demonstrate the superior performance of the MQI methodology, these results are based on the assumption that the underlying statistical model is correctly specified. Presumably, model misspecification would reduce the adequacy of MQI estimates. While investigating this issue is beyond the scope of this paper, the literature provides promising strategies to mitigate the risk of model misspecification, e.g., by the use of generalized additive mixed models (GAMM).\textsuperscript{15,26} Therefore, investigating effects of model misspecification and more flexible modeling strategies is an important task for future studies.

\clearpage

\hypertarget{references}{%
\section{References}\label{references}}

\hypertarget{refs}{}
\begin{CSLReferences}{0}{0}
\leavevmode\vadjust pre{\hypertarget{ref-AHRQ2022}{}}%
\CSLLeftMargin{1. }
\CSLRightInline{AHRQ. AHRQ quality indicators {[}Internet{]}. 2022. Available from: \url{https://qualityindicators.ahrq.gov/}}

\leavevmode\vadjust pre{\hypertarget{ref-CMS2022}{}}%
\CSLLeftMargin{2. }
\CSLRightInline{CMS. Hospital compare {[}Internet{]}. 2022. Available from: \url{https://www.cms.gov/Medicare/Quality-Initiatives-Patient-Assessment-Instruments/HospitalQualityInits/HospitalCompare}}

\leavevmode\vadjust pre{\hypertarget{ref-IQTIG2022}{}}%
\CSLLeftMargin{3. }
\CSLRightInline{IQTIG. Qualitätsindikatoren {[}Internet{]}. 2022. Available from: \url{https://iqtig.org/qs-instrumente/qualitaetsindikatoren/}}

\leavevmode\vadjust pre{\hypertarget{ref-DeLong1998}{}}%
\CSLLeftMargin{4. }
\CSLRightInline{DeLong ER, Peterson ED, DeLong DM, Muhlbaier LH, Hackett S, Mark DB. Comparing risk-adjustment methods for provider profiling. Statistics in Medicine {[}Internet{]}. 1998 Dec 4;16(23):2645--64. Available from: \url{http://dx.doi.org/10.1002/(SICI)1097-0258(19971215)16:23\%3C2645::AID-SIM696\%3E3.0.CO;2-D}}

\leavevmode\vadjust pre{\hypertarget{ref-Adams2017}{}}%
\CSLLeftMargin{5. }
\CSLRightInline{Adams M, Braun J, Bucher HU, Puhan MA, Bassler D, Von Wyl V. Comparison of three different methods for risk adjustment in neonatal medicine. BMC Pediatrics {[}Internet{]}. 2017 Apr 17;17(1). Available from: \url{http://dx.doi.org/10.1186/s12887-017-0861-5}}

\leavevmode\vadjust pre{\hypertarget{ref-Roessler2019}{}}%
\CSLLeftMargin{6. }
\CSLRightInline{Roessler M, Schmitt J, Schoffer O. Ranking hospitals when performance and risk factors are correlated: A simulation-based comparison of risk adjustment approaches for binary outcomes. PLOS ONE {[}Internet{]}. 2019 Dec;14(12):e0225844. Available from: \url{http://dx.doi.org/10.1371/journal.pone.0225844}}

\leavevmode\vadjust pre{\hypertarget{ref-Amin2019}{}}%
\CSLLeftMargin{7. }
\CSLRightInline{Amin R, Kitazawa T, Hatakeyama Y, Matsumoto K, Fujita S, Seto K, et al. {Trends in hospital standardized mortality ratios for stroke in Japan between 2012 and 2016: a retrospective observational study}. International Journal for Quality in Health Care {[}Internet{]}. 2019 Oct;31(9):G119--25. Available from: \url{https://doi.org/10.1093/intqhc/mzz091}}

\leavevmode\vadjust pre{\hypertarget{ref-Taylor2014}{}}%
\CSLLeftMargin{8. }
\CSLRightInline{Taylor P. {Standardized mortality ratios}. International Journal of Epidemiology {[}Internet{]}. 2014 Jan;42(6):1882--90. Available from: \url{https://doi.org/10.1093/ije/dyt209}}

\leavevmode\vadjust pre{\hypertarget{ref-Tambeur2019}{}}%
\CSLLeftMargin{9. }
\CSLRightInline{Tambeur W, Stijnen P, Vanden Boer G, Maertens P, Weltens C, Rademakers F, et al. Standardised mortality ratios as a user-friendly performance metric and trigger for quality improvement in a flemish hospital network: Multicentre retrospective study. BMJ Open {[}Internet{]}. 2019;9(9):e029857. Available from: \url{https://bmjopen.bmj.com/content/9/9/e029857}}

\leavevmode\vadjust pre{\hypertarget{ref-Yule1934}{}}%
\CSLLeftMargin{10. }
\CSLRightInline{Yule GU. On some points relating to vital statistics, more especially statistics of occupational mortality. Journal of the Royal Statistical Society {[}Internet{]}. 1934;97(1):1. Available from: \url{http://dx.doi.org/10.2307/2342014}}

\leavevmode\vadjust pre{\hypertarget{ref-Manktelow2014}{}}%
\CSLLeftMargin{11. }
\CSLRightInline{Manktelow BN, Evans TA, Draper ES. Differences in case-mix can influence the comparison of standardised mortality ratios even with optimal risk adjustment: An analysis of data from paediatric intensive care. BMJ Quality \& Safety {[}Internet{]}. 2014 Sep 1;23(9):782--8. Available from: \url{http://qualitysafety.bmj.com/content/23/9/782.abstract}}

\leavevmode\vadjust pre{\hypertarget{ref-Roessler2021}{}}%
\CSLLeftMargin{12. }
\CSLRightInline{Roessler M, Schmitt J, Schoffer O. Can we trust the standardized mortality ratio? A formal analysis and evaluation based on axiomatic requirements. PLOS ONE {[}Internet{]}. 2021;16(9):e0257003. Available from: \url{https://doi.org/10.1371/journal.pone.0257003}}

\leavevmode\vadjust pre{\hypertarget{ref-Rosenbaum1987}{}}%
\CSLLeftMargin{13. }
\CSLRightInline{Rosenbaum PR. Model-based direct adjustment. Journal of the American Statistical Association. 1987;82(398):387--94. }

\leavevmode\vadjust pre{\hypertarget{ref-Varewyck2016}{}}%
\CSLLeftMargin{14. }
\CSLRightInline{Varewyck M, Vansteelandt S, Eriksson M, Goetghebeur E. On the practice of ignoring center-patient interactions in evaluating hospital performance. Statistics in Medicine {[}Internet{]}. 2016;35(2):227--38. Available from: \url{https://onlinelibrary.wiley.com/doi/abs/10.1002/sim.6634}}

\leavevmode\vadjust pre{\hypertarget{ref-George2017}{}}%
\CSLLeftMargin{15. }
\CSLRightInline{George EI, Ročková V, Rosenbaum PR, Satopää VA, Silber JH. Mortality rate estimation and standardization for public reporting: Medicare{'}s hospital compare. Journal of the American Statistical Association {[}Internet{]}. 2017 Jul 3;112(519):933--47. Available from: \url{https://doi.org/10.1080/01621459.2016.1276021}}

\leavevmode\vadjust pre{\hypertarget{ref-Girling2012}{}}%
\CSLLeftMargin{16. }
\CSLRightInline{Girling AJ, Hofer TP, Wu J, Chilton PJ, Nicholl JP, Mohammed MA, et al. Case-mix adjusted hospital mortality is a poor proxy for preventable mortality: A modelling study. BMJ Quality \& Safety {[}Internet{]}. 2012;21(12):1052--6. Available from: \url{https://qualitysafety.bmj.com/content/21/12/1052}}

\leavevmode\vadjust pre{\hypertarget{ref-Lilford2010}{}}%
\CSLLeftMargin{17. }
\CSLRightInline{Lilford R, Pronovost P. Using hospital mortality rates to judge hospital performance: A bad idea that just won't go away. BMJ {[}Internet{]}. 2010;340. Available from: \url{https://www.bmj.com/content/340/bmj.c2016}}

\leavevmode\vadjust pre{\hypertarget{ref-Dishoeck2011}{}}%
\CSLLeftMargin{18. }
\CSLRightInline{Dishoeck AM van, Lingsma HF, Mackenbach JP, Steyerberg EW. Random variation and rankability of hospitals using outcome indicators. BMJ Quality \& Safety {[}Internet{]}. 2011;20(10):869--74. Available from: \url{https://qualitysafety.bmj.com/content/20/10/869}}

\leavevmode\vadjust pre{\hypertarget{ref-Silber2016}{}}%
\CSLLeftMargin{19. }
\CSLRightInline{Silber JH, Satopää VA, Mukherjee N, Rockova V, Wang W, Hill AS, et al. Improving medicare's hospital compare mortality model. Health Services Research {[}Internet{]}. 2016;51(S2):1229--47. Available from: \url{https://onlinelibrary.wiley.com/doi/abs/10.1111/1475-6773.12478}}

\leavevmode\vadjust pre{\hypertarget{ref-Davies2001}{}}%
\CSLLeftMargin{20. }
\CSLRightInline{Davies SM, Geppert J, McClellan M, McDonald K, Romano PS, Kaveh S. Refinement of the HCUP quality indicators. Agency for Healthcare Research; Quality; 2001. Report No.: 4. }

\leavevmode\vadjust pre{\hypertarget{ref-Marshall2006}{}}%
\CSLLeftMargin{21. }
\CSLRightInline{Marshall M, Klazinga N, Leatherman S, Hardy C, Bergmann E, Pisco L, et al. {OECD Health Care Quality Indicator Project. The expert panel on primary care prevention and health promotion}. International Journal for Quality in Health Care {[}Internet{]}. 2006 Sep;18(suppl\_1):21--5. Available from: \url{https://doi.org/10.1093/intqhc/mzl021}}

\leavevmode\vadjust pre{\hypertarget{ref-Elmen1996}{}}%
\CSLLeftMargin{22. }
\CSLRightInline{Elmén H, Höglund D, Karlberg P, Niklasson A, Nilsson W. {Birth weight for gestational age as a health indicator: Birth weight and mortality measures at the local area level}. European Journal of Public Health {[}Internet{]}. 1996 Jun;6(2):137--41. Available from: \url{https://doi.org/10.1093/eurpub/6.2.137}}

\leavevmode\vadjust pre{\hypertarget{ref-Snijders2011}{}}%
\CSLLeftMargin{23. }
\CSLRightInline{Snijders TA, Bosker RJ. Multilevel analysis: An introduction to basic and advanced multilevel modeling. 2nd ed. Sage; 2011. }

\leavevmode\vadjust pre{\hypertarget{ref-Krumholz2009}{}}%
\CSLLeftMargin{24. }
\CSLRightInline{Krumholz HM, Wang Y, Chen J, Drye EE, Spertus JA, Ross JS, et al. Reduction in Acute Myocardial Infarction Mortality in the United States. JAMA {[}Internet{]}. 2009 Aug 19;302(7):767. Available from: \url{http://dx.doi.org/10.1001/jama.2009.1178}}

\leavevmode\vadjust pre{\hypertarget{ref-Varewyck2014}{}}%
\CSLLeftMargin{25. }
\CSLRightInline{Varewyck M, Goetghebeur E, Eriksson M, Vansteelandt S. On shrinkage and model extrapolation in the evaluation of clinical center performance. Biostatistics {[}Internet{]}. 2014 May 8;15(4):651--64. Available from: \url{http://dx.doi.org/10.1093/biostatistics/kxu019}}

\leavevmode\vadjust pre{\hypertarget{ref-Gutzeit2022}{}}%
\CSLLeftMargin{26. }
\CSLRightInline{Gutzeit M, Rauh J, Kähler M, Cederbaum J. Modelling volume-outcome relationships in health care {[}Internet{]}. arXiv; 2022. Available from: \url{https://arxiv.org/abs/2203.12927}}

\end{CSLReferences}

\end{document}